\documentclass[prl,twocolumn,showpacs]{revtex4}
\usepackage{fancyhdr}
\usepackage{amssymb}
\usepackage{amsmath}
\usepackage[mathscr]{eucal}
\usepackage{latexsym}
\usepackage{amsbsy}
\usepackage{float}
\usepackage[dvips]{graphicx}
\usepackage{epsfig}
\usepackage{subfigure}
\usepackage{wrapfig}
\usepackage{appendix}

\newcommand{\be}{\begin{equation}}
\newcommand{\ee}{\end{equation}}
\newcommand{\bes}{\begin{equation*}}
\newcommand{\ees}{\end{equation*}}
\newcommand{\bea}{\begin{eqnarray}}
\newcommand{\eea}{\end{eqnarray}}
\newcommand{\bean}{\begin{eqnarray*}}
\newcommand{\eean}{\end{eqnarray*}}
\newcommand{\ba}{\begin{array}}
\newcommand{\ea}{\end{array}}
\newcommand{\tr}{\mathrm{tr}}
\newcommand{\Tr}{\mathrm{Tr}}

\newcommand{\Rv}{{\bf r}}
\newcommand{\bgamma}{\bar{\gamma}}
\newcommand{\hs}{\hspace{-0.045cm}}
\newcommand{\hsa}{\hspace{-0.10cm}}

\begin{document}

%\title{Enhancement of transition temperatures in fermionic systems with random hopping}
\title{Enhancement of critical temperatures in disordered bipartite lattices}

\author{Luca Dell'Anna}

\affiliation{Dipartimento di Fisica ed Astronomia ``Galileo Galilei'' and 
CNISM, Universit\`a di Padova, Italy}

\begin{abstract}
We study the strong enhancement, induced by random hopping, 
of the critical temperatures characterizing the transitions to 
superconductivity, charge-density wave and antiferromagnetism, 
which can occur in bipartite lattice models 
at half-filling, like graphene, by means of an extended 
Finkel'stein non-linear $\sigma$-model renormalization group approach. 
We show that, if Cooper channel interaction dominates,
 superconducting critical temperature can be enhanced at will, since
superconductivity cannot be broken by entering any Anderson insulating phase. 
%both in the presence of Anderson localization and in its absence, 
%close to the Gade-Wegner criticality. 
%In the latter case, which may be relevant for 
%electrons randomly hopping on a bipartite lattice at half-filling, 
If instead, staggered interactions are relevant, antiferromagnetic order 
%can be strongly promoted
is generated by disorder %and residual interactions 
at a temperature well above that expected for a clean system. 
\end{abstract}

\pacs{74.62.Dh, 71.30.+h, 72.15.Rn, 74.78.-w}

\maketitle

\section{Introduction}
\label{secIntro}
The interplay of disorder and interactions is the origin of several interesting and still unclear phenomena in condensed matter physics. 
One interesting problem raised in the past concerned the influence of 
randomness on Bardeen-Cooper-Schrieffer superconductors. 
It is well known that, in the absence of interaction, disorder can induce an 
insulating behavior on the electron systems, 
the so-called Anderson insulator \cite{anderson}. On the other hand, 
an attractive interaction makes the system unstable towards 
superconductivity. It was shown that a weak 
disorder does not spoil superconductivity \cite{anderson2} and the critical 
temperature $T_c$ is essentially unaffected by impurities \cite{lee, kapit}. 
However, in the presence of long-range Coulomb repulsion, diffusion of charges 
can lead to a suppression of superconducting $T_c$ \cite{fukuyama,finkelstein}. 
Quite recently, instead, it was shown that $T_c$
%the crtitical temperature of an s-wave superconductor 
can even be increased by Anderson localization \cite{kravtsov, kravtsov2, burmistrov}, 
provided that Coulomb interaction is screened and sufficiently weak. 

There are, however, disordered systems which do not show 
an Anderson insulating behavior, being close to the so-called Gade-Wegner 
criticality \cite{gade}. 
They are known as {\it two-sublattice models}, possessing a special symmetry 
called sublattice symmetry (namely, when only sites 
belonging to different sublattices are coupled), which usually describe 
particles randomly hopping (without on-site disorder) in nearest-neighbor 
sites 
on half-filled bipartite lattices, such as the square lattice or the honeycomb 
lattice, as in the case of graphene, which is naturally at half filling. 
Indeed when the impurity potential is close to the unitary scattering limit 
\cite{pepin} (when the impurity potential is infinitely strong) it reduces, by 
taking out one site, to a random nearest-neighbor hopping. This is what 
experimentally can be realized with graphene by 
substitutional doping or by vacancies. 
%Another possible mechanism which randomizes the hopping is given by static ripples present in graphene \cite{guinea}. Corrugations of the graphene sheet can naturally induce smooth fluctuations of the hopping parameter. 
The conductivity, with random hopping and in the absence of interactions,  
does not acquire any quantum interference corrections which would lead to 
Anderson localization, in contrast to systems with on-site disorder. 
The role of interactions in such systems, 
which are not Anderson insulators, is the missing piece of the puzzle. 

The important issue we will address in this paper is related to the question 
whether Cooper pair instability can be promoted by disorder in such systems 
and if disorder can unexpectedly improve a charge-density wave or 
generate a magnetic order. \\
We actually find that random hopping strongly enhances all the critical temperatures allowed in these systems, with respect to those obtained in the clean case, which delimit the transitions from normal phase to 
(i) superconductivity (SC), if the particle-particle Cooper channel dominates 
in the electron-electron interaction; 
(ii) charge-density wave (CDW), if, instead, a staggered particle-hole singlet 
channel is dominant; and, finally, (iii) antiferromagnet (AFM), 
if a staggered particle-hole triplet channel prevails. 
The main advantage of such random hopping two-sublattice systems as compared 
to standard systems (where sublattice symmetry is broken, for instance, 
by on-site disorder) is that one can improve almost {\it ad libitum} the 
transition temperatures, such as the superconducting critical temperature, 
by increasing the disorder parameters and tuning the residual 
interactions, never 
entering the Anderson insulating phase, which would break superconductivity.
In addition, other instabilities (AFM, CDW) 
are allowed, which are not present in the standard case. 
%
%are then very important to be understood. 
%interazione positiva -> SC
%disordine -> isolante di Anderson
%per piccolo disordine Anderson theorem-> SC persists
%dimostrato che Tc aumenta mentre con Coulomb diminuisce.
%sistemi disordinati non isolanti -> sistemi bipartiti (analogia al grafene)
%Qual e' il ground state? SC magnetic order. 
%

To study the role of randomness in electron systems, one can resort 
to a quantum field theory approach for disordered systems 
\cite{wegner,efetov}, further improved to deal with 
%to a well established theory describing 
combined effects of interactions and disorder. 
%in terms of a renormalized Fermi liquid 
\cite{finkelstein,castellani,belitz} 
%We will consider in particular systems close to half-filling, when the Mott insulating gap is hardly spoiled by randomness.
The interaction parameters acquire a scale dependence and, together with 
the conductance, form a full set of couplings of the so-called Finkel'stein 
non-linear $\sigma$ model, which flow under the action of the renormalization 
group (RG). 
Since we are interested in Cooper pair formations we will focus 
our attention to the systems in which time-reversal symmetry is preserved. 
The Wigner-Dyson class of symmetry covered by the standard Finkel'stein model 
is then the AI class (with time reversal and spin-rotation invariance, without 
sublattice symmetry). \cite{zirnbauer,mirlin,note} 
The symmetry class we are going to consider in this paper is the so-called 
BDI class, by the inclusion of the sublattice symmetry which produces 
anomalous behaviors already 
in the non-interacting case \cite{gade,michele}. 
One has, therefore, to extend the Finkel'stein model \cite{npb,ludwig} 
in the way shown in Appendix. 
%Supplemental Materials \cite{supplement}.
%we are going to shown in what follows. 

\section{RG equations for BDI class}
\label{secRGeq}
Considering, therefore, the case where 
%with the greatest symmetry, in which 
both sublattice symmetry and time-reversal symmetry are preserved (BDI class), 
%In this case, %which is the most symmetric one, 
%among those we shall be considering, 
%After defining, for convenience, the following parameters: 
%$g\equiv 1/(2\pi^2 \sigma)$, $\Gamma\equiv \Pi/(\sigma+n_r\Pi)$, 
%$\gamma^\alpha_s\equiv 2\nu\Gamma^\alpha_s/Z$, 
%$\gamma^\alpha_t\equiv 2\nu\Gamma^\alpha_t/Z$, 
%$\gamma^\alpha_c\equiv 2\nu\Gamma^\alpha_c/Z$, 
the complete %closed set of 
one-loop RG equations 
%for those couplings, in the limit $n_r\rightarrow 0$ and 
at $d=2+\epsilon$ dimensions, are given by Eqs. (\ref{(1)})-(\ref{(8)}). 
Since the dephasing scattering rate for our interacting particles
is basically given by the temperature $T$,
%being the coherence time for the quasiparticles,  
the integration of the RG equations will run from $T$ to
some energy cutoff
%$\tau^{-1}$, namely the elastic scattering rate.              
$\omega_o$, which, for our purposes, can be fixed by the Debye energy.
The scaling parameter is, therefore, given by $\ell=\ln(\omega_o/T)$. For
convenience we rescale $T\rightarrow \omega_o T$ so that the
diffusive regime we are going to consider is defined for $T\le 1$,
i.e. for $\ell=\ln(1/T)\ge 0$. Our bare starting parameters are taken
at $T= 1$, (at $\ell=0$), namely, at the scale corresponding to $\omega_o$.
%where $\ell=\ln(1/T)$ is the rescaling parameter. 
The equations are (see Ref. \cite{npb} for more details)
\begin{widetext}
\begin{eqnarray}
% (1)
\label{(1)}
\hspace{-0.3cm}\frac{d g}{d\ell} \hsa&=&\hsa -\epsilon g+
{g^2}\left\{6+\frac{1-\gamma_s^{0}}{\gamma_s^{0}}\,\ln(1-\gamma_s^{0})-\frac{1}{2}\gamma_s^{3}-3\,\frac{1+\gamma_t^{0}}{\gamma_t^{0}}\,\ln(1+\gamma_t^{0})
+\frac{3}{2}\gamma_t^{3}-\gamma_c^{0} 
%\right.\\&&\left. 
+ 2\,\frac{1-\gamma_c^{3}}{\gamma_c^{3}}\,\ln(1-\gamma_c^{3})\right\}
%\label{(1)}
\\
% (2)
\hspace{-0.3cm}\frac{d\Gamma}{d\ell}\hsa&=&\hsa
\epsilon \Gamma +8g +\frac{\Gamma}{g}\frac{dg}{d\ell}\\
% (3)   
\hspace{-0.3cm}\frac{d\gamma^{0}_t}{d\ell}\hsa&=&\hsa
g\left\{\left(1\hs+\hs\gamma_t^{0}\right)
\left(\hs-\frac{3}{2}\hs-\frac{\Gamma}{8}\hs+\hs\frac{\gamma_s^{0}}{2}\hs-\hs\frac{\gamma_s^{3}}{2}\hs-\hs\gamma_c^{0}\hs+\hs\gamma_c^{3}\hs+\hs\frac{3}{2}\gamma_t^{3}\right)
%\right.\\
%\nonumber &&\left.
\hs+\hs\left(1\hs+\hs\gamma_t^{0}\right)^2\left(\frac{3}{2}\hs+\hs\frac{\Gamma}{8}\hs+\hs\gamma_s^{3}\hs-\hs\gamma_t^{3}\Big(1\hs-\hs\frac{\Gamma}{4}\Big)\hs+\hs2\gamma_c^{0}\right)\hs-\hs 2(\gamma_t^{3})^2
%\gamma_t^{3}
\right\}\\
% (4)                  
\label{(4)}
\hspace{-0.3cm}\frac{d\gamma^{3}_t}{d\ell}\hsa&=&\hsa g\left\{
\left(\frac{1}{2}+\frac{\Gamma}{4}\right)\gamma_t^{0}+\left(\frac{3}{2}+\frac{\Gamma}{8}+\ln\left[\frac{(1-\gamma_c^{3})^2(1-\gamma_s^{^{0}})}{1+\gamma_t^{0}}\right]\right)
\gamma_t^{3}+
\frac{\gamma_s^{3}}{2}+\frac{\gamma_s^{0}}{2}+\gamma_c^{0}+\gamma_c^{3}\right.\\
\nonumber&&+\gamma_t^{3}\left(\frac{1}{2}\gamma_s^{0}-\frac{3}{2}\gamma_t^{0}+\gamma_c^{3}
\left.+\gamma_c^{0}-\frac{3}{2}\gamma_t^{3}+\frac{1}{2}\gamma_s^{3}\right)\right\}+(\gamma_t^{3})^2\\
% (5)
\hspace{-0.3cm}\frac{d\gamma^{0}_s}{d\ell}\hsa&=&\hsa
g\left\{\left(1\hs-\hs\gamma_s^{0}\right)
\left(\frac{1}{2}\hs+\hs\frac{\Gamma}{8}\hs+\hs\frac{\gamma_s^{3}}{2}\hs+\hs\gamma_c^{0}\hs-\hs\gamma_c^{3}\hs+\hs\frac{3}{2}(\gamma_t^{0}\hs-\hs\gamma_t^{3})\right)
%\right.\\
%\nonumber&&\left.
\hs-\hs \left(1\hs-\hs\gamma_s^{0}\right)^2\left(\frac{1}{2}\hs+\hs\frac{\Gamma}{8}\hs+\hs\gamma_s^{3}\Big(1\hs-\hs\frac{\Gamma}{4}\Big)\hs-\hs\frac{2\gamma_c^{3}}{1-\gamma_c^{3}}
-3\gamma_t^{3}\right)\hs+\hs2(\gamma_c^{0})^2
%\gamma_c^{0}
\right\}\phantom{,}\\
% (6)    
\label{(6)}    
\hspace{-0.3cm}\frac{d\gamma^{3}_s}{d\ell}\hsa&=&\hsa g\left\{
\frac{3}{2}\gamma_t^{0}+\frac{3}{2}\gamma_t^{3}+\left(\frac{1}{2}+\frac{\Gamma}{8}+\ln\left[(1+\gamma_t^{0})^3(1-\gamma_s^{0})(1-\gamma_c^{3})^2\right]\right)
\gamma_s^{3}-\left(\frac{1}{2}-\frac{\Gamma}{4}\right)\gamma_s^{0}+\gamma_c^{3}\right.\\
\nonumber&&+\gamma_c^{0}\left(1-4\ln(1-\gamma_c^{3})\right)
+\gamma_s^{3}\left(\frac{1}{2}\gamma_s^{0}-\frac{3}{2}\gamma_t^{0}+\gamma_c^{3}
\left.-\frac{3}{2}\gamma_t^{3}+\frac{1}{2}\gamma_s^{3}+\gamma_c^{0}\right)\right\}-(\gamma_s^{3})^2 \phantom{--}\\
% (7)    
\label{(7)}     
\hspace{-0.3cm}\frac{d\gamma^{0}_c}{d\ell}\hsa&=&\hsa g\left\{
\frac{3}{2}\gamma_t^{0}+\frac{3}{2}\gamma_t^{3}+\frac{\gamma_s^{0}}{2}+\frac{\gamma_s^{3}}{2}+\left(1+\frac{\Gamma}{8}+
\ln\left[\frac{(1+\gamma_t^{0})^3(1-\gamma_c^{3})^2}{(1-\gamma_s^{0})}\right]\right)\gamma_c^{0}+\frac{\Gamma}{4}\gamma_c^{3}\right.\\
\nonumber&&-2\gamma_s^{3}\ln(1-\gamma_c^{3})
+\gamma_c^{0}\left(\frac{1}{2}\gamma_s^{0}-\frac{3}{2}\gamma_t^{0}+\gamma_c^{3}
\left.-\frac{3}{2}\gamma_t^{3}+\frac{1}{2}\gamma_s^{3}+\gamma_c^{0}\right)\right\}-(\gamma_c^{0})^2\\
% (8)      
\label{(8)}    
\hspace{-0.3cm}\frac{d\gamma^{3}_c}{d\ell}\hsa&=&\hsa g\left\{-1-\gamma_c^{0}+\gamma_s^{0}+\gamma_s^{3}
-2\gamma_c^{3}\left(\ln(1-\gamma_c^{3})-\frac{\gamma_c^{3}}{\gamma_s^{0}}\ln(1-\gamma_s^{0})\right)
+2\gamma_c^{0}\gamma_s^{3}\right.\\
\nonumber&&+\left(1-\gamma_c^{3}\right)\left(3+\frac{\Gamma}{8}+3\gamma_c^{0}-\frac{\gamma_s^{0}}{2}-\frac{3}{2}\gamma_s^{3}
+\frac{3}{2}\left(\gamma_t^{0}-\gamma_t^{3}\right)\right)
%\\&&
\left.+\left(1-\gamma_c^{3}\right)^2\left(-2-\frac{\Gamma}{8}-2\gamma_c^{0}+\frac{\Gamma}{4}\gamma_c^{0}+\gamma_s^{3}+3\gamma_t^{3}\right)\right\}
%\label{(8)}
%%%%%%%%%%%%%%%%%%%%%%%%%%%%%%%%%%%%%%%%%
\end{eqnarray}
\end{widetext}
In Eqs.~(\ref{(1)}-\ref{(8)}) the disorder parameters are: 
$g$, the charge resistivity, and $\Gamma$, related to the non-zero mean bond 
dimerization of the original lattice \cite{gade,michele,npb,ludwig}. 
Notice that, for $\epsilon=0$ and in the limit of all $\gamma\rightarrow 0$,  
Eq.~(\ref{(1)}) becomes $dg/d\ell=0$, namely, $g$ remains constant. 
This non-interacting behavior is what is called the Gade-Wegner criticality \cite{gade,mirlin}. 
The interaction parameters are related to 
(i) %small momentum transfer 
smooth interactions: $\gamma_s^0$ (particle-hole singlet channel), $\gamma_t^0$ (particle-hole triplet channel), $\gamma_c^0$ (particle-particle Cooper channel); and (ii) staggered sublattice interactions: $\gamma_s^3$ (particle-hole singlet), $\gamma_t^3$ (particle-hole triplet), and $\gamma_c^3$ (particle-particle).
Among these parameters, $\gamma_t^3$, $\gamma_s^3$ and $\gamma_c^0$ are 
responsible for antiferromagnetic spin density wave (AFM), 
charge density wave (CDW) and $s$-wave superconductivity (SC), respectively.
The very last terms in Eqs.~(\ref{(4)}), (\ref{(6)}, and (\ref{(7)}), those 
not coupled to $g$, are actually, the terms 
which can drive the system 
to AFM ($\gamma_t^3$), CDW ($\gamma_s^3$), and SC ($\gamma_c^0$) also 
in clean systems, obtained by simple ladder summations. Equations 
(\ref{(1)})-(\ref{(8)}) are quite complicated; nevertheless, they hide 
an amazing property: they, in fact, are symmetric under the transformation 
$\gamma_s^0=\gamma_c^3\leftrightarrow -\gamma^0_t$ and  
$\gamma_s^3=\gamma_c^0\leftrightarrow -\gamma^3_t$. 
This symmetry property of the parameters can be obtained by 
particle-hole transformation of the original fermionic fields defined on the lattice, $c_{i\uparrow} \rightarrow c_{i\uparrow}$, $c_{i\downarrow} \rightarrow (-)^ic^\dagger_{i\downarrow}$, which maps charge to spin and vice versa.
By imposing %, in the peculiar particle-hole symmetric case,
%such a transformation that the full system does not change under particle-hole transformation we derive the following equivalences among interaction couplings
%\bea
%\label{gamma}
$\gamma^0_s=\gamma_c^3=-\gamma_t^0\equiv\gamma $, and 
$\gamma^3_s=\gamma_c^0=-\gamma_t^3\equiv\bgamma$, 
%\label{bgamma}
%\eea
namely, considering the peculiar particle-hole symmetric case, 
%which surprisingly simplify and reduce 
Eqs.~(\ref{(1)})-(\ref{(8)}) are strongly simplified (see Appendix)
%\cite{supplement}. %as follows
%\begin{eqnarray}
%\label{(11)}
%&&\hspace{-0.3cm}\frac{dg}{d\ell}=-\epsilon g+3g^2\left\{2\left(1+\frac{1-\gamma}{\gamma}\ln(1-\gamma)\right)-\bgamma\right\}\\
%&&\hspace{-0.3cm}\frac{d\Gamma}{d\ell}=\epsilon \Gamma +8g +\frac{\Gamma}{g}\frac{dg}{d\ell}\\
%&&\hspace{-0.3cm}\frac{d\gamma}{d\ell}=g\left\{\left(1-\gamma\right)\left(\frac{\Gamma}{8}+3\bgamma\right) +2\bgamma^2\right.\\
%&&\hspace{-0.3cm}\nonumber\phantom{d\gamma}\left.+(1-\gamma)^2\left(\frac{\Gamma}{4}\bgamma-4\bgamma-\frac{\Gamma}{8}\right)\right\}\\
%\label{(14)}
%&&\hspace{-0.3cm}\frac{d\bgamma}{d\ell}=g\left\{
%\bgamma\left(\frac{\Gamma}{8}+2\ln(1-\gamma)+3(\gamma+\bgamma)\right)\right.\\
%&&\hspace{-0.3cm}\nonumber \phantom{d\bgamma}\left.+\left(\frac{\Gamma}{4}-1\right)\gamma\right\}-\bgamma^2
%\end{eqnarray}
%We have found, in other words, that 
%Eqs.~(\ref{gamma}-\ref{bgamma}), actually, 
These conditions, actually,  
define a subspace of the full space 
of parameters, invariant under RG flow. 
%which is invariant under RG flow. 
%namely, if the starting point of the RG is chosen into that subspace, then the RG flow will never escape.

%Solving these sets of equations we found that we get instabilities for temperatures well above the critical temperatures one can obtain in the clean systems.

\section{Analytical solutions}
\label{secAn}
%Supposing $\Gamma(\ell=0)\approx 0$, 
For starting amplitudes fulfilling 
$|\gamma_c^0|\gg |\gamma_s^{0}|, |\gamma_t^0|, |\gamma_c^3|, |\gamma_t^3|, |\gamma_s^3|$, 
%and for $|\gamma_c^0|\gg|\gamma_t^3|,\,|\gamma_s^3|$ 
Eqs.~(\ref{(1)})-(\ref{(8)}) can be approximated as follows: 
$dg/d\ell\simeq -\epsilon g$, 
$d\Gamma/d\ell \simeq 8 g$, 
$d\gamma_s^0/d\ell\simeq d\gamma_t^0/d\ell\simeq d\gamma_s^3/d\ell\simeq d\gamma_t^3/d\ell\simeq g\gamma_c^0$, 
$d\gamma_c^3/d\ell\simeq \Gamma g\gamma_c^0/4$, 
and finally
\be
\label{eq_gamc}
\frac{d\gamma_c^0}{d\ell}\simeq \eta\,%g\left(1+\frac{\Gamma}{8}\right)
\gamma_c^0-(\gamma_c^0)^2\,,
\ee
%\bea
%
%\eea
where $\eta\equiv z-d =g\left(1+\frac{\Gamma}{8}\right)$, with $z$ being the dynamical exponent. 
The solution of Eq.~(\ref{eq_gamc}), for $\epsilon=0$, is 
\be
\label{gammac(l)}
%\gamma_c^0(\ell)\simeq \gamma_{c}^0\frac{e^{\frac{1}{2}\left(\frac{\Gamma(\ell)}{8}\right)^2}}{1-\frac{|\gamma_{c}^0|}{g_0}\sqrt{\frac{{\pi}}{{2}}}\,\textrm{erfi}\left(\frac{\Gamma(\ell)}{8\sqrt{2}}\right)}
%
\hspace{-0.05cm}
\gamma_c^0(\ell)\simeq \frac{\gamma_{c}^0\,e^{
%1+\frac{\Gamma(\ell)}{8}
\frac{\eta(\ell)^2}{2g_0^2}}}{e^{\frac{\eta_0^2}{2g_0^2}}
+\frac{\gamma_{c}^0}{g_0}\sqrt{\frac{\pi}{2}}\left(
\textrm{erfi}\left(
%\frac{\Gamma(\ell)+8}{8\sqrt{2}}
\frac{\eta(\ell)}{\sqrt{2}g_0}\right)-
\textrm{erfi}\left(\frac{\eta_0}{\sqrt{2}g_0}\right)\right)}
\ee
where $\gamma_c^0$ here means $\gamma_c^0(0)$, to make notation simpler, 
%$\gamma_c^0=\gamma_c^0(0)$, is the starting value, 
$\textrm{erfi}(y)=\frac{2}{\sqrt{\pi}}\int_0^y e^{x^2}dx$ is the imaginary 
error function and $\eta(\ell)\simeq \eta_0+g_0^2\ell$, with $g_0=g(0)$ and 
$\eta_0=g_0(1+\frac{\Gamma_0}{8})$.\\
%$\Gamma(\ell)\simeq 8g_0\ell$.
%For simplicity, let us assume $\Gamma_0\equiv \Gamma(0)\simeq 0$, so that $\eta_0\simeq g_0$. 
For $\gamma_c^0<0$ the system is unstable towards superconductivity whose $T_c$, for $\eta_0\ll|\gamma_c^0|$, at leading orders, is 
%one finds the superconducting critical temperature %by solving 
%\be
%\label{inst}
%%\frac{g_0}{|\gamma^0_c|}
%%\simeq\sqrt{\frac{{\pi}}{{2}}}\,\textrm{erfi}\left(\frac{g_0 \ln(1/T_c)}{\sqrt{2}}\right)
%\frac{g_0}{|\gamma^0_c|}  
%\simeq\sqrt{\frac{{\pi}}{{2e}}}
%\left(\textrm{erfi}\left(\frac{1-g_0 \,\ln T_c}{\sqrt{2}}\right)-
%\textrm{erfi}\left(\frac{1}{\sqrt{2}}\right)
%\right)
%\ee
%which, for $g_0\ll|\gamma_c^0|$, gives
%\be
%\label{Tc_gsmall} 
\be
\label{Tc_gsmall}
T_c\sim 
\exp\left(-\frac{1}{|\gamma_c^0|}\left(1-\frac{\eta_0}{2|\gamma_c^0|}\right)\right).
\ee
%at leading orders.
%after neglecting terms $O(({g_0}/{|\gamma_c^0|})^2)$ in the exponent. 
One can see that, for $g_0\rightarrow 0$, namely for a clean system, one recovers 
the known result $T_c=T_c^{BCS}\sim e^{-{1}/{|\gamma_c^0|}}$. In other words, the disorder parameter $\eta_0$, which is always positive, improves $T_c$. The 
enhancement is stronger for large disorder. 
For $\eta_0\sim g_0\gg|\gamma_c^0|$, in fact, %from Eq.~(\ref{inst}) 
we get
\be
\label{Tc_appx}
T_c\sim \exp\left(-\frac{{\cal C}%(g_0,\gamma_c^0)
}{g_0}\right)\,,
\ee
%\phantom{..}\\
where  
${\cal C}={\cal C}(g_0,\gamma_c^0)\approx \sqrt{W\left((g_0/|\gamma_c^0|)^2\right)}$, a smooth function ($W(x)$ is the Lambert function or product logarithm).
%g_0/|\gamma_c^0|-\frac{1}{2}(g_0/\gamma_c^0)^2$, for $g_0\ll |\gamma_c^0|$, such that, in this limit, 
%%\exp[{-1/|\gamma_c^0|+g_0/(2|\gamma_c^0|^2)}] 
%$T_c\simeq T_c^{BCS} \exp\left[{g_0}/{(2|\gamma_c^0|^2)}\right] $, therefore, in the clean limit we recover $T_c^{BCS}\simeq \exp[-1/|\gamma_c^0|]$.
%For $g_0\gg |\gamma_c^0|$, instead, ${\cal F}(g_0/\gamma_c^0)\approx (2 \ln(g_0/|\gamma_c^0|))^{1/4}$, namely, is a slow function of $g_0$.

%In the same way one can study the instabilities towards spin or charge 
%density waves when only $\gamma_t^3$ or $\gamma_s^3$ dominate, finding 
%analogous expressions for the corresponding critical temperatures, 
%where $\gamma_c^0$ is replaced by $\gamma_t^3$ or $\gamma_s^3$. 
In the special 
%case described by 
%Eqs.~(\ref{gamma}-\ref{bgamma}), namely in the 
particle-hole symmetric case, 
%the equations for small $\gamma$, are 
%$d\gamma/d \ell \simeq g\left((\Gamma/4-1)\bar\gamma+\gamma\Gamma/8\right)$ 
%and 
%$d\bar\gamma/d\ell\simeq g\left(\bar\gamma\Gamma/8+(\Gamma/4-1)\gamma\right)-(\bar\gamma)^2$. 
%%Just neglecting the term proportional to $\gamma$ in the last equation, and for $\Gamma_0\simeq 0$, we get  
%%$
%%\bar\gamma(\ell)\simeq \bar\gamma
%%\frac{e^{\frac{1}{2}\left(\frac{\Gamma(\ell)}{8}\right)^2}}
%%{1-\frac{|\bar\gamma|}{g_0}
%%\sqrt{\frac{\pi}{2}}\textrm{erfi}\left(\frac{\Gamma(\ell)}{8\sqrt{2}}\right)}
%%$
%%which, 
%%Neglecting $\gamma$, for $\Gamma_0\simeq 0$ and 
for $g_0\ll|\bar\gamma|$ and $\Gamma_0\simeq 0$, %neglecting $\gamma$, %yields 
%we get the following critical temperature 
%$T_c\sim \exp\left(-\frac{1}{|\bar\gamma|}\left(1-\frac{g_0^2}{6|\bar\gamma|^2}\right)\right)$, namely, 
$\ln T_c$ increases quadratically with $g_0$ (see Appendix) 
%\cite{supplement}, 
instead of linearly as in Eq.~(\ref{Tc_gsmall}). %seen before. 
This deviation from linearity is observed already when $|\gamma_s^3|$ and $|\gamma_t^3|$ become of the same order of $|\gamma_c^0|$, as shown in Fig.~\ref{fig.sc-sdw}, for small $g_0$. 
%in Eq.~(\ref{Tc_gsmall}). 
%For sufficiently strong disorder, $g_0\gg |\bar \gamma|$, the 
%critical temperature goes like Eq.~(\ref{Tc_appx}), after replacing 
%$\gamma_c^0$ with $\bar\gamma$. It is worth stressing that, 
%%under the peculiar condition dictated by 
%in the particle-hole symmetric case, Eqs.~(\ref{gamma}-\ref{bgamma}), 
%%superconductivity, charge and spin density waves 
%AFM, CDW and SC occur simultaneously. 
%%exhibiting a strange phase that we can call {\it super-spin-solid}.

Before concluding this section, a couple of comments is in order. 
As declared also in Ref. \cite{burmistrov}, strictly in 2D the 
SC transition is of the 
Berezinskii-Kosterliz-Thouless (BKT) type, whereas we have calculated the 
mean-field transition temperature (which identifies  
Cooper pairs formation). However, since the mean-field and 
BKT temperatures do not differ much \cite{beasley}, 
we expect that the enhancement of $T_c$ holds also for $T_{BKT}$. 
Finally, for very strong disorder ($g\,\Gamma\gg 1$) we would expect that
the dynamical exponent is affected by a sort of 
electron freezing effect \cite{motrunich,mudry,npb1}, 
a weak multifractal effect which 
modifies $z$ as $z\sim \sqrt{g\,\Gamma}$. 
As a result, in the strong disorder regime, 
Eq.~(\ref{Tc_appx}) should turn into  
$T_c\sim \exp\left(-{\tilde{\cal C}   %(g_0,\gamma_c^0)
}/{g^{{2}/{3}}}\right)$. %with $C^\prime$ another function. 
%instead of Eq.~(\ref{Tc_appx}). 

For $\epsilon>0$, since $\frac{d\eta}{d\ell}=-\epsilon \eta+O(g^2)$, Eq.~(\ref{eq_gamc}), can be rewritten as 
$
\frac{d\gamma_c^0}{d\eta}\simeq -\frac{\gamma_c^0}{\epsilon}+\frac{(\gamma_c^0)^2}{\epsilon \eta}
$
whose solution, for $\eta_0\ll \epsilon$, gives the following 
critical temperature 
%$T_c\sim\exp\left(-\frac{1}{|\gamma_c^0|}\left(1-\frac{\eta_0}{\epsilon}\right)\right)$, 
\be
\label{Tc-epsilon}
T_c\sim \left(T_c^{BCS}\right)^{1-({\eta_0}/{\epsilon})}. 
%T_c\sim (T_c^{BCS})^{e^{-\eta_0/\epsilon}}
\ee
$T_c$ is still enhanced by disorder with respect to 
%the critical temperature of a clean system, 
$T_c^{BCS}$. 
Notice that the disorder parameter, $\eta_0\sim g_0$, can be arbitrarily 
strong, compared to $|\gamma_c^0|$, 
%so that both the temperatures 
and that $T_c$ and $T_c^{BCS}$ $<1$, being 
both rescaled by the 
%Debye 
energy cutoff $\omega_o$.
%($T_c\sim (T_c^{BCS})^{(1-\eta_0/\epsilon))}$)

Analogously, one can study the instabilities towards spin or charge density
waves when only $\gamma_t^3$ or $\gamma_s^3$ dominate, finding similar 
expressions for the corresponding critical temperatures. 
%where $\gamma_c^0$ is replaced by $\gamma_t^3$ or $\gamma_s^3$. 
In particular for $|\gamma_s^3|$ greater than all the other parameters, 
%\gg |\gamma_s^{0}|, |\gamma_t^0|,|\gamma_c^3|, |\gamma_t^3|, |\gamma_c^0|$ 
we have to solve
\be
\frac{d\gamma_s^3}{d\ell}\simeq \left(\eta-\frac{g}{2}\right)\,\gamma_s^3-(\gamma_s^3)^2
%\;\;\;\;\textrm{for}\;|\gamma_s^3|\gg |\gamma_s^{0}|, |\gamma_t^0|, |\gamma_c^3|, |\gamma_t^3|, |\gamma_c^0|\\
\ee
whose solution, for $\gamma_s^3<0$, gives a critical temperature for CDW with the same 
behaviors as in Eqs.~(\ref{Tc_gsmall})-(\ref{Tc-epsilon}) 
with $\gamma_c^0$ replaced 
by $\gamma_s^3$, $\eta_0$ by $(\eta_0-g_0/2)$, and $T_c^{BCS}$ by $\exp{(-1/|\gamma_s^3|)}$. For $|\gamma_t^3|$ dominant, instead, we have  
\be
\frac{d\gamma_t^3}{d\ell}\simeq \left(\eta+\frac{g}{2}\right)\,\gamma_t^3+(\gamma_t^3)^2
%\;\;\;\;\textrm{for}\;|\gamma_t^3|\gg |\gamma_s^{0}|, |\gamma_t^0|, |\gamma_c^3|, |\gamma_s^3|, |\gamma_c^0|\\
\ee
which, for $\gamma_t^3>0$, drives the system to AFM with a new N\'eel 
temperature given by Eqs.~(\ref{Tc_gsmall})-(\ref{Tc-epsilon}), 
where $\gamma_c^0$ is replaced by $\gamma_t^3$, $\eta_0$ by $(\eta_0+g_0/2)$, 
and $T_c^{BCS}$ by the N\'eel temperature $\exp{(-1/\gamma_t^3)}$ of the clean system.
\section{RG solutions}
\label{secRG}

\begin{figure}[ht]
\includegraphics[width=8.cm]{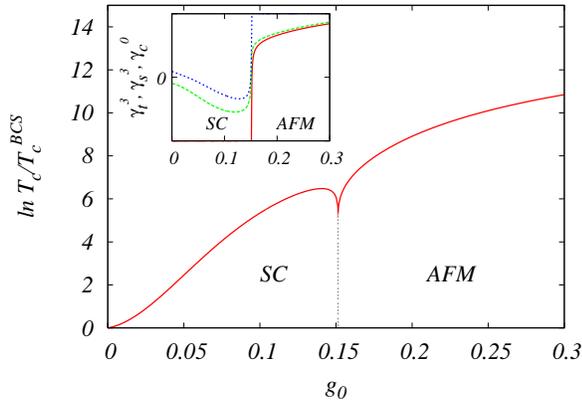}%{Tc_SC-SDW_new.eps}
\caption{(Color online) Critical temperature, in log-scale, in units of $T_c^{BCS}$
%(critical temperature for clean superconductor) 
as a function of the initial 
value for $g$ ($g_0\equiv g(\ell=0)$), obtained for the following starting
parameters: $\epsilon=0$, $\Gamma(\ell=0)=0$, $\gamma_s^0(0)=\gamma_t^0(0)=\gamma_c^3(0)
=0.01$,
$\gamma_c^0(0)=-0.065$, $\gamma_s^3(0)=-0.04$, $\gamma_t^3(0)=0.04$. For
$g_0<0.15$, $T_c$ is the superconducting critical temperature 
(SC) while for $g_0>0.15$, $T_c$ is the N\'eel temperature for
antiferromagnets (AFM). 
The vertical dotted line only projects $T_c$ where the two instabilities
exchange. Inset: the parameters, $\gamma_s^3$ (green long dashed line), $\gamma_t^3$
(blue dashed line),
$\gamma_c^0$ (red solid line), at $T=T_c$, as functions of $g_0$.}
\label{fig.sc-sdw}
\end{figure}
\begin{figure}[ht]
\includegraphics[width=8.cm]{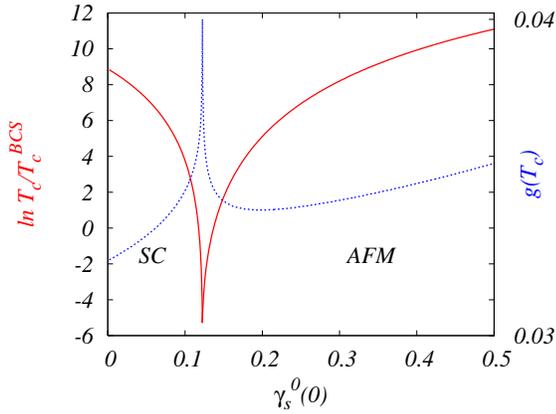}%{Tc_s_BDI_4.eps}
\caption{(Color online) Critical temperature (red-solid line), in log-scale,
in units of $T_c^{BCS}$
%(critical temperature for clean superconductor)
as a function of the initial
value for the particle-hole singlet interaction $\gamma_s^0(\ell=0)$, at $\epsilon=0$, obtained for the following initial 
parameters: $g_0=0.03$, $\Gamma(0)=0$, $\gamma_t^0(0)=-\gamma_c^3(0)=0.001$,
$\gamma_c^0(0)=-0.04$, $\gamma_t^3(0)=-\gamma_s^3(0)=0.001$. For $\gamma_s^0(0)<0.12245$, $T_c$ is the SC critical temperature, otherwise it is the critical temperature for AFM. For the same initial conditions, 
the renormalized $g$, when $T=T_c$, is plotted (blue-dashed line) as a function of $\gamma_s^0(0)$.}
%In the inset the instabilities parameters, $\gamma_s^3$ (green long dashed line), $\gamma_t^3$ (blue dashed line), $\gamma_c^0$ (red solid line), close to $T_c$, as functions of $g_0$.}           
\label{fig.sc-sdw-gammas}
\end{figure}
A richer variety of behaviors can be found by 
solving the full set of Eqs.~(\ref{(1)})-(\ref{(8)}) (by using the FORTRAN 
code provided in Ref. \cite{supplement}). 
It can happen that small and large disorder regimes can be characterized by 
different phases.  
%we get a richer variety of behaviors. %\cite{suplement}. 
In particular, we found that AFM is the most favoured instability, provided 
that $\gamma_t^3>0$. If $\gamma_c^0$ is the dominant parameter but 
$\gamma_t^3$ is also sizable, by increasing disorder ($g_0$) the 
superconducting $T_c$ is enhanced up to a value of $g_0$ above which 
AFM may prevail, whose corresponding critical temperature is 
much higher than the N\'eel temperature in a clean system, 
i.e. $T_c\gg \exp(-1/\gamma_t^3)$ (see Fig.~\ref{fig.sc-sdw}). 
An important role in the occurrence of AFM is played 
also by the other parameters, even when they start with small values. It 
is crucial, therefore, to take into account all the contributions 
appearing in Eqs.~(\ref{(1)}-\ref{(8)}), 
in order to draw correctly the boundaries of different phases. 
%SC and AFM domains. 
By fixing the disorder strength $g_0$ but increasing the bare parameter 
$\gamma_s^0(0)$, the 
SC $T_c$ is suppressed. For sufficiently large $\gamma_s^0$, then the dominant 
instability turns to be the AFM again (see Fig.~\ref{fig.sc-sdw-gammas}). 
In other words, 
we can go from an $s$-wave superconducting regime to a magnetic ordered 
regime, not only by simply increasing the triplet staggered interaction, 
but also by increasing disorder or increasing the singlet slow repulsive 
interaction. In all these cases the N\'eel temperature $T_c$ is strongly 
enhanced by the presence of disorder. Moreover the critical interaction for 
getting an antiferromagnet can be tuned by disorder strength. 
This last result can be relevant in graphene 
 where an antiferromagnetic order is believed to occur above some critical 
interaction and where random nearest-neighbor hopping can be mimicked by 
substitutional doping or vacancies.

\begin{figure}[ht]
\includegraphics[width=8cm]{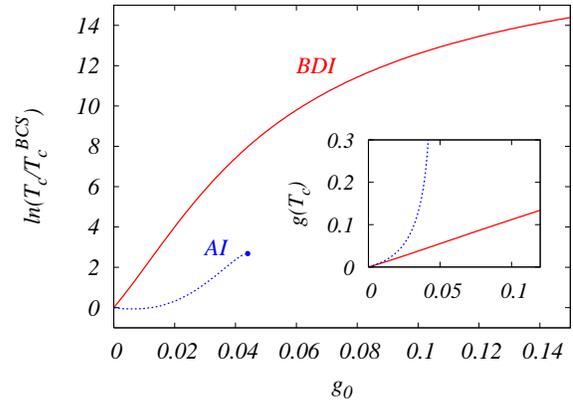}%{Tc_s_g0.03c0.04.eps}          
\caption{(Color on line) Comparison of SC $T_c$ for BDI and AI classes.  
%as functions of $g_0$. 
In both cases all the initial
parameters are zero except $\gamma_c^0(0)=-0.05$. The increase of 
$T_c$ in the AI case (blue-dashed line, in agreement with Ref. \cite{burmistrov}) stops entering the Anderson 
insulating regime (where $g$ diverges, see the corresponding blue-dashed line in the inset) while $T_c$ in the BDI case keeps on increasing with $g_0$ (red solid line).}
\label{fig.fink-bdi}                                                           
\end{figure}  

\section{Conclusions}
\label{secCon}

Interestingly, we have found a strong enhancement of the superconducting critical temperature, Eq.~(\ref{Tc_appx}), for BDI class of 
two-sublattice models in two-dimensions (like a honeycomb), usually characterized by 
electron systems at 
half filling with a random hopping, which are not Anderson insulators. 
%and Eq.~(\ref{finkTc}) 
$T_c$
can be even larger than that obtained in the %does not differ much from the 
AI standard case for Anderson insulators \cite{burmistrov} 
(see Appendix and Fig.~\ref{fig.fink-bdi}), 
usually obtained by on-site disorder, going out of half-filling or 
breaking somehow the sublattice symmetry.  
%even though the two systems are quite different. 
In the BDI case, the dominant 
corrections to $T_c$ are given by the rescaling of the dynamical exponent (or, 
in other words, %the renormalization 
of the density of states). 
On the contrary, in the AI case, 
%where the compressibility is not renormalized \cite{finkelstein}, 
$z=d$, therefore $\eta=0$, and  the increase of $T_c$
%Eq.~(\ref{finkTc}) 
originates from an equation similar to Eq.~(\ref{eq_gamc}), 
where $\eta$ is replaced by $g$, strongly 
renormalized by Anderson quantum interference corrections. 
Finally for a 
system at the Anderson transition (two- and three-dimensional (3D) 
%2D and 3D 
symplectic class and 3D orthogonal class) $\eta$ is replaced by the fractal exponent $d_2$, getting a 
power-law enhancement of $T_c$ \cite{kravtsov}. 
The importance of studying the two-sublattice systems 
%definitely relevant for graphene, 
%in addition to what done so far, 
is also the appearance of other instabilities such as charge-density wave 
or antiferromagnetic order induced by disorder.  
%which can be definitely relevant for graphene. 
Moreover, the great advantage of such systems with random hopping, 
belonging to BDI class, 
as compared to the standard case with on-site disorder, is that 
the superconducting $T_c$ can be increased at will (at least within 
the validity of one-loop calculation, i.e. for $g_0\lesssim 1$) by tuning 
disorder and 
residual repulsive interaction, since superconductivity cannot be broken  
by entering any Anderson insulating phase (see Fig.~\ref{fig.fink-bdi}). 
\section{Acknowledgements} 
\label{sec}
The author acknowledges financial support from MIUR (``ArtiQuS'', FIRB 2012 RBFR12NLNA\_$002$), thanks SISSA, Trieste, for hospitality and A. Trombettoni for useful discussions.

\section{Appendix}
\appendix
\section{Extended Finkel'stein non-linear $\sigma$ model}
The effective low-energy model which describe transverse charge fluctuations, 
derived in analogy with the original Finkel'stein model \cite{finkelstein}, is 
the following \cite{npb}
\be
S=S_{0}+S_{I}\,,
\label{act}
\ee
where $S_0$ is the non-interacting part
\bea
\nonumber
S_{0}&=&\frac{\pi}{32}
\int d\Rv\, \left\{\sigma\, \Tr\left(\vec{\nabla}Q \cdot\vec{\nabla} 
Q^\dagger\right)\right.
-{8\nu Z}\,
\Tr\left(\hat\omega Q\right)\\
\label{NLsM}
&&
\left.
-\frac{\Pi}{8}\,
\Tr\left(Q^\dagger\vec{\nabla}Q\rho_3\right)
\cdot \Tr\left(Q^\dagger\vec{\nabla}Q\rho_3\right)
\right\},
\label{NLsM}
\eea
and $S_I$ the contribution from $e$-$e$ interactions
\bea
\nonumber S_{I}=\frac{\pi^2\nu^2}{32}\hspace{-0.1cm}
\int^{\prime}
\hspace{-0.1cm}
\Big\{
\Gamma^{\alpha}_t
\sum_{\beta=0,3}\tr(Q^{ii}_{n,n+p} \vec{S}_{\alpha\beta})
\cdot\tr(Q^{ii}_{\ell+p,\ell}\vec{S}_{\alpha\beta})\\
\nonumber
-\Gamma^{\alpha}_s
\sum_{\beta=0,3}\tr(Q^{ii}_{n,n+p} S^0_{\alpha\beta})
\,\tr(Q^{ii}_{\ell+p,\ell}S^0_{\alpha\beta}) \\
+\Gamma^{\alpha}_c
\sum_{\beta=1,2}\tr(Q^{ii}_{n+p,-n} S^0_{\alpha\beta})
\,\tr(Q^{ii}_{\ell+p,-\ell} S^0_{\alpha\beta}
\Big\}
\label{inter}
\eea
The symbol $\int^\prime$ in Eq.~(\ref{inter}) means an integral over real
space, a sum over smooth ($\alpha=0$) and staggered
($\alpha=3$) modes, $n_r$ replica indices and
Matsubara frequencies, i.e.
$\int^\prime\equiv \int d\Rv\sum_{\alpha=0,3}\sum_i                        
\sum_{\ell,n,p}$, where $i$ is the replica index and $\ell, n, p$ are
Matsubara indices.
The matrix field $Q$ is constrained by the condition $QQ^\dagger=\mathbb{I}$.
The coupling $\sigma$
corresponds to the Kubo formula for the charge conductivity at the Born level;
$\nu$ is the density of states at the Fermi energy at the Born approximation;
$\Gamma^0_s, \Gamma^0_t, \Gamma^0_c$ and
$\Gamma^3_s, \Gamma^3_t, \Gamma^3_c$ are related to the Landau scattering 
amplitudes \cite{belitz} or to the interaction parameters of a bipartite 
Hubbard-like model \cite{ludwig} ($-t\sum c^\dagger_{Ai}c_{Bj}+h.c.
+U \sum(n_{Ai}n_{Ai'}+n_{Bj}n_{Bj'})+V \sum n_{Ai} n_{Bj}$),
for smooth ($\Gamma^0\sim(U+V)$) and 
staggered sublattice ($\Gamma^3\sim(U-V)$) components, 
in the particle-hole singlet, particle-hole triplet 
and particle-particle Cooper channels, respectively; $Z$ is the field
renormalization constant; $\hat\omega$ is a diagonal matrix made of Matsubara
frequencies. The last term in Eq.~(\ref{NLsM}) is the anomalous additional
term which is present only if the sublattice symmetry is preserved
\cite{gade} and the coupling $\Pi$ is
related to the staggered density of states fluctuations \cite{michele}.
$\vec{S}_{\alpha\beta}$ is a vector made of three tensor products, i.e. 
$\vec{S}_{\alpha\beta}=\rho_\alpha\otimes \tau_\beta\otimes \vec{\sigma}$, 
while $S^0_{\alpha\beta}=\rho_\alpha\otimes \tau_\beta\otimes \sigma_0$ is a 
single tensor; $\tau_1, \tau_2, \tau_3$ are Pauli matrices in particle-hole space;
$\sigma_1, \sigma_2, \sigma_3$ are Pauli matrices in spin space; $\rho_3$ is 
the third Pauli matrix in the sublattice space;
$\tau_0, \sigma_0, \rho_0$ are identity matrices in the corresponding spaces.
The trace ``$\Tr$'' is made over all spaces (particle-hole, spin, sublattice,
replica and Matsubara spaces), while the trace ``$\tr$''
is over particle-hole, spin and sublattice spaces.
The action in Eqs.~(\ref{act})-(\ref{inter}) is tailored to describe
two-sublattice models, namely when sublattice symmetry is preserved and
staggered modes are massless.
When the sublattice symmetry is broken, staggered modes become massive, then
the last term in
Eq.~(\ref{NLsM}) %should be removed                               
and the terms in Eq.~(\ref{inter}) with
$\alpha=3$ should be put to zero ($\Pi=\Gamma_s^3=\Gamma_t^3=\Gamma_c^3=0$).
In this way one recovers the standard
Finkel'stein non-linear $\sigma$-model \cite{finkelstein}, provided that
$Q$ takes values in the proper coset space.
If, instead, staggered modes are massless, those interacting terms are
naturally generated by the RG flow.\\
In order to get rid of the field renormalization parameter, it is usually
convenient to define the following parameters:
$g\equiv 1/(2\pi^2 \sigma)$, $\Gamma\equiv \Pi/(\sigma+n_r\Pi)$,
$\gamma^\alpha_s\equiv 2\nu\Gamma^\alpha_s/Z$,
$\gamma^\alpha_t\equiv 2\nu\Gamma^\alpha_t/Z$,
$\gamma^\alpha_c\equiv 2\nu\Gamma^\alpha_c/Z$.

\section{RG equations - BDI class in the particle-hole symmetric case}
In the presence of sublattice symmetry and imposing the particle-hole
invariance by
\bea
\label{gamma}
\gamma^0_s=\gamma_c^3=-\gamma_t^0\equiv\gamma \\
\gamma^3_s=\gamma_c^0=-\gamma_t^3\equiv\bgamma
\label{bgamma}
\eea
Eqs. (1-8) reported in the Letter reduce simply to
\begin{eqnarray}
\label{(11)}
&&\hspace{-0.3cm}\frac{dg}{d\ell}=-\epsilon g+3g^2\left\{2\left(1+\frac{1-\gamma}{\gamma}\ln(1-\gamma)\right)-\bgamma\right\}\\
&&\hspace{-0.3cm}\frac{d\Gamma}{d\ell}=\epsilon \Gamma +8g +\frac{\Gamma}{g}\frac{dg}{d\ell}\\
&&\hspace{-0.3cm}\frac{d\gamma}{d\ell}=g\left\{\left(1-\gamma\right)\left(\frac{\Gamma}{8}+3\bgamma\right) +2\bgamma^2\right.\\
&&\hspace{-0.3cm}\nonumber\phantom{d\gamma}\left.+(1-\gamma)^2\left(\frac{\Gamma}{4}\bgamma-4\bgamma-\frac{\Gamma}{8}\right)\right\}\\
\label{(14)}
&&\hspace{-0.3cm}\frac{d\bgamma}{d\ell}=g\left\{
\bgamma\left(\frac{\Gamma}{8}+2\ln(1-\gamma)+3(\gamma+\bgamma)\right)\right.\\
&&\hspace{-0.3cm}\nonumber \phantom{d\bgamma}\left.+\left(\frac{\Gamma}{4}-1\right)\gamma\right\}-\bgamma^2
\end{eqnarray}
For $|\gamma|\ll|\bar\gamma|$ we have
$d\gamma/d \ell \simeq g\left((\Gamma/4-1)\bar\gamma+\gamma\Gamma/8\right)$
and
$d\bar\gamma/d\ell\simeq g\left(\bar\gamma\Gamma/8+(\Gamma/4-1)\gamma\right)-
(\bar\gamma)^2$.
For $g_0\ll|\bar\gamma|$ and $\Gamma_0\simeq 0$, neglecting $\gamma$, 
we get the following critical temperature
\be
T_c\sim \exp\left(-\frac{1}{|\bar\gamma|}\left(1-\frac{g_0^2}{6|\bar\gamma|^2}
\right)\right),
\ee
namely, $\ln T_c$ increases quadratically with $g_0$.
For sufficiently strong disorder, $g_0\gg |\bar \gamma|$, the
critical temperature goes like
$T_c\sim \exp\left(-\frac{{\cal C}%(g_0,\gamma_c^0)                  
}{g_0}\right)$, as in the general case treated in the Letter.
It is worth stressing that,
in the particle-hole symmetric case fixed by Eqs.~(\ref{gamma}-\ref{bgamma}),
AFM, CDW and SC occur simultaneously.

\subsection{RG equation in the standard case - AI class}
\begin{figure}[ht]
\includegraphics[width=8cm]{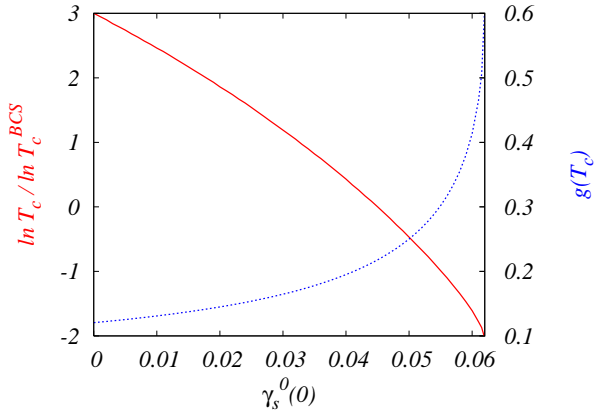}%{Tc_s_g0.03c0.04.eps}                     
\caption{(Color online) Critical temperature (red-solid line), in log-scale, 
for AI class, in units of $T_c^{BCS}$
%(critical temperature for clean superconductor)                             
as a function of the initial
value for $\gamma_s^0(0)$, at $\epsilon=0$, obtained for the following starting
parameters: $g_0=0.03$, $\gamma_t^0(0)=0$, and
$\gamma_c^0=-0.04$. For the same parameters
$g(T=T_c)$ is plotted (blue dashed line) as a function of
the starting values $\gamma_s^0(0)$.}
\label{fig.fink-sc-gammas}
\end{figure}

If we break the sublattice symmetry, by going far from half-filling or in the 
presence of on-site impurities,
we obtain the one-loop Finkel'stein equations, not restricted to long range 
Coulomb case \cite{finkelstein}, at all orders in the interaction strenghts,
\begin{widetext}
\begin{eqnarray}
\label{(f1)}
&&\hspace{-0.cm}\frac{dg}{d\ell}=-\epsilon g+g^2\left\{5+
\frac{1-\gamma_s^{0}}{\gamma_s^{0}}\,\ln(1-\gamma_s^{0})
-3\,\frac{1+\gamma_t^{0}}{\gamma_t^{0}}\,\ln(1+\gamma_t^{0})-\gamma_c^{0}
\right\}\\
&&\frac{d\gamma_s^{0}}{d\ell}=g\left\{\left(1-\gamma_s^{0}\right)\left
(\frac{3}{2}\gamma_t^{0}+\gamma_c^{0}-\frac{1}{2}\right)
+\frac{1}{2}\left(1-\gamma_s^{0}\right)^2+2(\gamma_c^{0})^2\right\}\\
&&\frac{d\gamma_t^{0}}{d\ell}=g\left\{\left(1+\gamma_t^{0}\right)\left(\frac{1}{2}\gamma_s^{0}-\gamma_c^{0}-\frac{1}{2}\right)
+\left(1+\gamma_t^{0}\right)^2\left(\frac{1}{2}+2\gamma_c^{0}\right)\right\}\\
\label{(f4)}
&&\frac{d\gamma_c^{0}}{d\ell}=g\left\{\frac{\gamma_s^{0}}{2}+\frac{3}{2}
\gamma_t^{0}
+\gamma_c^{0}\left(\frac{\gamma_s^{0}}{2}-\frac{3}{2}\gamma_t^{0}+\gamma_c^{0}+
\ln\left[\frac{(1+\gamma_t^{0})^3}{(1-\gamma_s^{0})}\right]\right)\right\}-
\left(\gamma_c^{0}\right)^2
\end{eqnarray}
\end{widetext}
which reduce to the equations reported in Ref. \cite{burmistrov}, if only 
first orders in the $\gamma$'s are considered. In this case, for $\epsilon=0$, 
and $g_0\gg |\gamma_c^0(0)|$, one recovers the result \cite{burmistrov}
\be
\label{finkTc}
T_c\sim \exp{(-1/g_0)}\,.
\ee
Solving now the full set of Eqs.~(\ref{(f1)}-\ref{(f4)}), we can show
that, under increasing the bare repulsive interaction, $\gamma_s^0(\ell=0)$,
$T_c$ decreases, and calculate the values of $\gamma_s^0(0)$ for
which $T_c>T_c^{BCS}$ or $T_c<T_c^{BCS}$,
as shown by Fig.~\ref{fig.fink-sc-gammas}.


\begin{thebibliography}{99}
\bibitem{anderson} P.W. Anderson, Phys. Rev. {\bf 109}, 1492 (1958).
\bibitem{anderson2} P.W. Anderson, J. Phys. Chem. Solids {\bf 11}, 26 (1959).
\bibitem{lee} M. Ma and P.A. Lee, Phys. Rev. B {\bf 32}, 5658 (1985).
\bibitem{kapit} A. Kapitulnik and G. Kotliar, Phys. Rev. Lett. {\bf 54}, 473 (1985).
\bibitem{fukuyama} S. Maekawa, H. Fukuyama,J. Phys. Soc. Jpn. {\bf 51} 1380 (1982).
\bibitem{finkelstein} A.M. Finkel'shtein, Zh. Eksp. Teor. Fiz. {\bf 84} 168 (1983); Sov. Phys. JETP {\bf 57} 97 (1983); A.M. Finkel'stein, Z. Phys. B {\bf 56} 189 (1984).
\bibitem{kravtsov} M.V. Feigelman, L.B. Ioffe, V.E. Kravtsov, E.A. Yuzbashyan, Phys. Rev. Lett. {\bf 98} 027001 (2007). 
\bibitem{kravtsov2} M.V. Feigelman, L.B. Ioffe, V.E. Kravtsov, E. Cuevas, Ann. Phys. {\bf 325} 1390 (2010).
\bibitem{burmistrov} I.S. Burmistrov, I.V. Gornyi, A.D. Mirlin, Phys. Rev. Lett. {\bf 108}, 017002 (2012)
\bibitem{gade} R. Gade, F. Wegner, Nucl. Phys. B {\bf 360} 213 (1991); R. Gade, Nucl. Phys. B {\bf 398} 499 (1993).
\bibitem{pepin} C. Pepin, P.A. Lee, Phys. Rev. Lett. {\bf 81}, 2779 (1998).
%\bibitem{guinea} F. Guinea, M.I. Katsnelson, M.A.H. Vozmediano, Phys. Rev. B {\bf 77}, 075422 (2008).
\bibitem{wegner} F.J. Wegner Z. Phys. B {\bf 35}, 207 (1979).
\bibitem{efetov} K.B. Efetov, A.I.Larkin, and D.E. Khmel'nitskii, Zh. Eksp. Teor. Fiz. {\bf 79}, 1120 (1979); Sov. Phys. JETP {\bf 52}, 568 (1980).
\bibitem{castellani} C. Castellani, C. Di Castro, P.A. Lee, M. Ma, S. Sorella, E. Tabet, Phys. Rev. B {\bf 30}, 1596 (1984).
\bibitem{belitz} D. Belitz, T.R. Kirkpatrick, Rev. Mod. Phys. {\bf 66}, 
261 (1994).
\bibitem{zirnbauer} M. R. Zirnbauer, J. Math. Phys. {\bf 37}, 4986 (1996).
\bibitem{mirlin} F. Evers, Mirlin, Rev. Mod. Phys. {\bf 80}, 1355 (2008).
\bibitem{note} In the symmetry classification of disordered systems 
the Cartan symbol AI refers to GOE (Gaussian orthogonal ensemble) 
random matices and the $\sigma$-model target space is $Sp(2n)/Sp(n)\times Sp(n)$ (in the replica formulation, $n\rightarrow 0$). BDI is the chiral class for GOE random matrices, in the $\sigma$-model formulation it 
corresponds to the manifold $U(2n)/Sp(n)$. 
\bibitem{michele} M. Fabrizio, C. Castellani, Nucl. Phys. B {\bf 583} 542 (2000).
\bibitem{npb} L. Dell'Anna, Nucl. Phys. B {\bf 758} 255 (2006).
\bibitem{ludwig} M.S. Foster and A.W.W. Ludwig, Phys. Rev. B {\bf 77}, 165108 
(2008).
%\bibitem{supplement} See Supplemental Materials for the extended Finkel'stein non linear $\sigma$-model and the RG equations for AI class and BDI particle-hole symmetric case. Fortran code solving RG equations for both BDI and AI cases is also provided.
\bibitem{beasley} M.R. Beasley, J.E. Mooij, and T.P. Orlando, Phys. Rev. Lett. {\bf 42}, 1165 (1979).
\bibitem{motrunich} O. Motrunich, K. Damle, and D.A. Huse, Phys. Rev. 
B {\bf 65}, 064206 (2002).
\bibitem{mudry} C. Mudry, S. Ryu, and A. Furusaki, Phys. Rev. B {\bf 67},
  064202 (2003). 
\bibitem{npb1} L. Dell'Anna, Nucl. Phys. B {\bf 750} 213 (2006).
\bibitem{supplement} Find at \href{http://arxiv.org/format/1306.4618}{http://arxiv.org/format/1306.4618} the source file of the FORTRAN code (RG.f) solving the RG equations for both BDI, Eqs.~(\ref{(1)})-(\ref{(8)}), and AI, Eqs.~(\ref{(f1)})-(\ref{(f4)}), cases.  
%in Supplemental Materials.
\end{thebibliography}
\end{document}